\documentclass[12pt,thmsa]{article}
\usepackage{sw20bams}

\input tcilatex
\QQQ{Language}{
American English
}

\begin{document}

\title{Synchronization and Inertial Frames}
\author{C. Viazminsky \\
%EndAName
International Institute of Theoretical and Applied Physics,\\
Iowa State University, Ames, IA 50011, USA\\
and Department of Physics, University of Aleppo, Syria}
\maketitle

\begin{abstract}
In classical mechanics a procedure for simultaneous synchronization in all
inertial frames is consistent with the Galilean transformation. However, if
one attempts to achieve such synchronization utilizing light signals, then
he will be facing on one hand a break down of absolute simultaneity, and on
the other hand, a self-contradictory transformation that has Lorentz
transformation, or the confinement of Lorentz transformation to the velocity
of light, as the only possible ways that resolve the contradiction. The
current work constitutes a smooth transition from traditional to
relativistic vision of mechanics, and therefore is quite appealing from
pedagogical point of view.
\end{abstract}

\section{Introduction}

Time in Newtonian mechanics is considered as an absolute entity, so that one
can choose an instant $t_0$ which can be taken as the initial time in all
inertial frames, and such that when an event occurs at an instant $t$ in one
inertial frame then it occurs also at the same instant $t$ in all inertial
frames. The inquiry of how, in principle, can one synchronize the clocks in
all inertial frames receives a number of plausible answers. One possible
hypothetical procedure is as follows: it is sufficient to synchronize clocks
in one inertial frame, say $S$, with a master clock at some point in $S$,
say the origin $O$, using a stream of particles emitted radially from $O$
with the same velocity $C$, so that on first receiving the stream, an
observer at a point $P$ sets his clock on $t=t_0+\left| \stackrel{%
\rightarrow }{r}\right| /C,$ where $\stackrel{\rightarrow }{r}$ is the
position vector of the point $P$ and $t_0$ is the time read by the master
clock when the stream starts\cite{Rindler}. In this way the space is
furnished with a set of synchronized clocks which can be adopted in all
inertial frames. In fact it is straight forward to see, using the Galilean
transformation for the coordinates and the velocity, that the procedure we
have just described does synchronize also the clocks in every other inertial
frame with the master clock at the origin of $S$, and accordingly with each
other. However if it turns out that the law of velocity addition in Galilean
transformation is not applicable to the stream of particles then
synchronization is beyond reach because the concept of absolute simultaneity
ceases to be valid.

We present here a procedure for synchronization using a stream of photons,
and show that in spite of one's attempt to adhere to the familiar concepts
of Newton mechanics, the shortcomings of these concepts are revealed when
synchronization is attempted for more than one inertial frame. We shall
adopt from start the experimental fact \cite{Fillipas}that the velocity of
light is independent of the state of relative motion between the source and
the observer, which is equivalent to say that this velocity is a constant $C$%
, which is the same in all inertial frames.

\section{Synchronization by Light Signals}

Let $S$ and $S^{\prime }$ be two inertial frames comprising two systems of
rectangular Cartesian coordinates $oxyz$ and $o^{\prime }x^{\prime
}y^{\prime }z^{\prime }$ respectively, in standard configuration. Assume
that $S^{\prime }$ is moving relative to $S$ in the direction of the $x$%
-axis with a uniform velocity $u$ $(u>0)$. When the plane $x^{\prime }=0$ in 
$S^{\prime }$ coincide with the plane x$=0$ in $S$, every observer in the
plane $x^{\prime }=0$ sets his clock $T^{\prime }=0$ and simultaneously
emits a pulse of light of short duration parallel to the $x^{\prime }-$axis.
The assembly of these pulses generates a plane wave propagating in the $\pm
x^{\prime }$ directions with velocity $C$. The furthest wave front appears
to the observer $O^{\prime }$ at an instant $T^{\prime }$ as occupying two
planes $\left| x^{\prime }\right| =CT^{\prime },$ and hence all observers in
these two planes have to set their clocks at $T^{\prime }.$ Since time flaws
uniformly in $S^{\prime },$ all observers in the region \{X$^{\prime
}:\left| X^{\prime }\right| \leq CT^{\prime }\}$ have the same clock
reading, namely $T^{\prime }.$

For the $S$-observer the plane wave we have just prescribed appears as
originating from the plane $x=0$ and travelling in the $\pm x-$directions
with velocity $C$, for the light velocity has been assumed to be constant
and independent of the source's motion. The observer $O$ uses the Galilean
transformation to deduce that the equation of the light's front is 
\begin{equation}
x=x^{^{\prime }}+uT^{^{\prime }},  \label{e1}
\end{equation}
which represents two planes corresponding to $x^{\prime }=\pm CT^{\prime }$
in $S^{\prime }.$

An $S$-observer at a location $(x,y,z)$, on first receiving the wave has to
set his clock at $T=\left| x\right| /C.$ Dividing (1) by $C$ and making use
of $T^{\prime }=\left| x^{\prime }\right| /C$ we get the equation 
\begin{equation}
T=T^{\prime }+ux^{\prime }/C^2,  \label{e2}
\end{equation}
which relates the initial time settings in both frames for the points lying
on the wave front. Although at an instant $T^{\prime }$ in $S^{\prime },$
all clocks in the region $\left| X^{\prime }\right| \leq CT^{^{\prime
}}=\left| x^{^{\prime }}\right| $ record the same time $T^{\prime },$ the
corresponding clocks in $S$ in the region $-x^{^{\prime }\prime }+uT^{\prime
}\leq X\leq \left| x^{\prime }\right| +uT^{\prime },$ or equivalently $%
(-C+u)T^{\prime }\leq X\leq (C+u)T^{\prime },$ do not read the same time. In
fact all those corresponding to $x^{\prime }=CT^{^{\prime }}$ read $%
T=T^{\prime }(1+u/C),$ whereas those corresponding to $x^{\prime
}=-CT^{\prime }$ read $T=T^{\prime }(1-u/C).$ In other words time is no more
absolute in $S$, i.e. in the system which performed synchronization using
the Galilean transformation, and at the same time accepting that the
velocity of light waves emitted from the plane $x^{\prime }=0$ in $S^{\prime
}$ is also equal to $C$ in $S$. The peculiar result concerning time in $S$
is expected, because the velocity in classical mechanics cannot be
independent of the inertial frame relative to which it is measured.

\section{Lorentz Transformation}

In matrix notation we write the transformation given by (1,2), together with
the obviously valid relations $y=y^{\prime },z=z^{\prime }$ as 
\begin{equation}
(Txyz\widetilde{)}=A(T^{\prime }x^{\prime }y^{\prime }z^{\prime }\widetilde{)%
}  \label{e3}
\end{equation}
where the symbol ($\widetilde{})$ denoted the transpose of a four-vector,
and $A$ is the transformation matrix 
\begin{equation}
A=\left[ 
\begin{array}{cccc}
1 & u/C^2 & 0 & 0 \\ 
u & 1 & 0 & 0 \\ 
0 & 0 & 1 & 0 \\ 
0 & 0 & 0 & 1
\end{array}
\right]  \label{e4}
\end{equation}
On physical grounds, and on interchanging the roles of the $S$ and $%
S^{\prime }-$observers, we find that the transformation from $S$ to $%
S^{\prime }:$%
\begin{equation}
(T^{\prime }x^{\prime }y^{\prime }z^{\prime }\widetilde{)}=A(Txyz\widetilde{)%
},  \label{e5}
\end{equation}
results through replacing $u$ in $A$ by $(-u)$ to obtain 
\begin{equation}
A^{\prime }=\left[ 
\begin{array}{cccc}
1 & -u/C^2 & 0 & 0 \\ 
-u & 1 & 0 & 0 \\ 
0 & 0 & 1 & 0 \\ 
0 & 0 & 0 & 1
\end{array}
\right] .  \label{e6}
\end{equation}
The composite transformation (3) and (5), i.e. from $S^{\prime }$ to $S$ and
back to $S^{\prime }$ must certainly be the identity transformation. We have
however 
\begin{equation}
AA^{\prime }=\left[ 
\begin{array}{cc}
(1-u^2/C^2)I_2 & 0 \\ 
0 & I_2
\end{array}
\right] ,  \label{e7}
\end{equation}
which is a contradiction. The way out of this contradiction is to scale the
right hand-sides of the expressions of $T$ and $x$ as given by (1) and (2),
and also the similar expressions of $T^{\prime }$ and $x^{\prime },$ through
multiplication by $\gamma =(1-u^2/C^2)^{-1/2},$ to obtain 
\begin{eqnarray}
x &=&\frac{x^{\prime }+uT^{^{\prime }}}{\sqrt{1-u^2/C^2}}\;,\;\;y=y^{\prime
},\;\;z=z^{\prime }  \label{e8} \\
T &=&\frac{T^{\prime }+ux^{\prime }}{\sqrt{1-u^2/C^2}}\;.  \nonumber
\end{eqnarray}

Lorentz transformation is obtained on postulating that the last relations
are valid for an arbitrary moment $t^{\prime }$ in $S^{\prime },$ i.e. we
have to replace $T^{\prime }$ in (8) by an arbitrary $t^{\prime }$ and the
corresponding $T$ by $t$. For more details concerning Lorentz transformation
and their properties we refer to \cite{Born,French,Landau,Lawden}. As far as
our work is concerned it is important to note that Lorentz transformation is
a point transformation in a four dimensional manifold $(t,x,y,z)$, in the
sense that each point (event) $(t,x,y,z)$ in $S$ is described by a unique
point $(t^{\prime },x^{\prime },y^{\prime },z^{\prime })$ in $S^{\prime },$
and hence the classical vision of absolute time is no more valid. Therefore
our goal of achieving simultaneous synchronization of inertial frames is
proved impossible, and instead, every frame is synchronized independently.
For more details about the new aspects of space and time we refer to
reference\cite{Davies}.

\section{An Alternative Result}

For the points of the synchronization wave front, the relations 
\begin{equation}
\left| x\right| =CT,\;\;\;\left| x^{\prime }\right| =CT^{\prime }  \label{e9}
\end{equation}
are valid. Hence we may write (1) and (2), according to $x^{\prime }$ being
positive or negative in either form

(i) For $x^{\prime }>0$%
\begin{equation}
x=(1+u/C)x^{\prime },\;\;\;\;\;T=(1+u/C)T^{\prime }.  \label{e10}
\end{equation}

(ii) For $x^{\prime }<0$%
\begin{equation}
x=(1-u/C)x^{\prime },\;\;\;\;T=(1-u/C)T^{\prime }.\;\;  \label{e11}
\end{equation}
Since $u/C<1$, it is evident that $x$ and $x^{\prime }$ are both positive or
both negative. An identical statement is also valid for $T$ and $T^{\prime
}. $ It is evident, in both cases (i) and (ii), that the two equations
obtained are linearly dependent, and therefore are reducible to one
equation. In fact the second equation in each case is obtained through
dividing both sides of the first equation by $C$ or $(-C)$. We shall
therefore confine our attention to the second equation in each case.

Due to symmetry the transformation from $S$ to $S^{\prime }$ is obtained by
replacing $u$ by $(-u)$ in (10) and (11); and hence must be

(i) For $x>0$ 
\begin{equation}
T^{\prime }=(1-u/C)T  \label{e12}
\end{equation}

(ii) For $x<0$ 
\begin{equation}
T^{\prime }=(1+u/C)  \label{e13}
\end{equation}
But these are not the inverse transformation as deduced from (10) and (11).
In fact by (10-13) we have the contradiction:

For $\;x>0$ \ \ \ \ \ \ \ \ \ \ \ \ \ \ \ $T=(1+u/C)T^{\prime }=(1-u^2/C^2)T$

For $\;x<0$\ \ \ \ \ \ \ \ \ \ \ \ $T=(1-u/C)T^{\prime }=(1-u^2/C^2)T,$

which is resolved as before by multiplying the right hand-sides of the
relations (10-13) by the factor $\gamma =(1-u^2/C^2)^{-1/2}$ to obtain 
\begin{equation}
x=ax^{\prime },\;\;\;\;\;T=aT^{\prime }\;\;\;\;\;(x^{\prime }>0),
\label{e14}
\end{equation}
\begin{equation}
x=x^{\prime }/a,\;\;\;\;T=T^{\prime }/a\;\;\;\;(x^{\prime }<0),  \label{e15}
\end{equation}
where 
\begin{equation}
a=\sqrt{(C+u)/(C-u)}.  \label{e16}
\end{equation}
Replacing $u$ by $(-u)$ gives the inverse transformation 
\begin{equation}
x^{\prime }=x/a,\;\;\;\;\;T^{\prime }=T/a\;\;\;\;\;(x>0),  \label{e17}
\end{equation}
\begin{equation}
x^{\prime }=ax,\;\;\;\;\;\;T^{\prime }=aT\;\;\;\;\;(x<0).  \label{e18}
\end{equation}
The transformation (14,15) or the inverse transformation (17,18) gives the
relations between the coordinates of an arbitrary point in both frames the
moment it receives the wave of synchronization, or equivalently, the
relation between the time settings by the $S$ and $S^{\prime }$-observers at
that point when they first receive the wave.

It important to notice that the transformation we have just found may be
deduced from Lorentz transformation. Actually if we set $\left| x^{\prime
}\right| =CT^{\prime }$ in (8) we get the transformation (14,15). Note that
for $\left| x^{\prime }\right| =CT^{\prime }$ which is the equation of the
fronts of two waves originating simultaneously from the plane $x^{\prime }=0$
and travelling in opposite directions, are transformed by Lorentz
transformation (8) into two planes given by (14,15). The last two planes are
not symmetric with respect to $x=0,$ because the simultaneous events ($%
T^{\prime },x=\pm CT^{\prime },y^{\prime },z^{\prime })$ in $S^{\prime }$
are not simultaneous in $S$.

Perhaps it is worthy to mention that the relations (14,15) are equivalent to
the special Lorentz transformation (8) in which we still have the
constraints $\left| x^{\prime }\right| =CT^{\prime },$ and which is
applicable only to the positions occupied by the light front and the
corresponding times. Lorentz transformation results indeed from (8) only
when we abandon this constraint and deal with $x^{\prime }$ and $T^{\prime }$
as independent variables.

Our approach therefore, leads either to Lorentz transformation itself, or to
Lorentz transformation applicable only to the light propagation. Only a
daring step in which the coordinates and time are liberated from the
constraint, we have mentioned, leads to Lorentz transformation.

\section{Conclusion}

The Galilean transformation is completely compatible with the concept of
absolute time in classical mechanics. However, when one attempts to
incorporate the source-independent constant velocity of light in Galilean
transformation then he will be faced from start by a type of transformation,
namely the wrong intermediate transformation (1,2) which terminates the
concept of simultaneity and accordingly absolute time. Following up this
result and accepting the symmetric roles of inertial frames we deduce that
the transformation from $S$ to $S^{\prime }$ result from (1,2) by simply
replacing $u$ by $(-u)$. The final transformation we are seeking must be
satisfied by the coordinates of any event that are measured by observers in
the two frames. The mathematical expression of the last statement is that
the transformation from one frame to another must be inverse to each other.
This requirement determines a factor $(\gamma )$ by which we have to scale
the right hand-sides of the intermediate transformation to obtain formally
the Lorentz transformation. Although, it is known to every physicist that
Lorentz transformation is the correct substitute of the Galilean
transformation, the method we have followed to derive this transformation is
new and has nice features manifested in the smooth transition from Newtonian
to relativistic concepts.

\end{document}